\documentclass[aps,pra,reprint,superscriptaddress]{revtex4-1}

\usepackage{graphicx}
\usepackage{multirow}
\usepackage{amsmath}

\begin{document}

\title[]{Lamb-dips and Lamb-peaks in the saturation spectrum of HD}%

\author{M.L. Diouf}
\affiliation{Department of Physics and Astronomy, LaserLaB, Vrije Universiteit, De Boelelaan 1081, 1081 HV Amsterdam, The Netherlands}

\author{F.M.J. Cozijn}
\affiliation{Department of Physics and Astronomy, LaserLaB, Vrije Universiteit, De Boelelaan 1081, 1081 HV Amsterdam, The Netherlands}

\author{B. Darqui\'e}
\affiliation{Laboratoire de Physique des Lasers, Universit\'e Paris 13, Sorbonne Paris Cit\'e et CNRS, 99 av J B Cl\'ement, 93430 Villetaneuse, France}

\author{E.J. Salumbides}
\affiliation{Department of Physics and Astronomy, LaserLaB, Vrije Universiteit, De Boelelaan 1081, 1081 HV Amsterdam, The Netherlands}

\author{W. Ubachs}
\email{w.m.g.ubachs@vu.nl}
\affiliation{Department of Physics and Astronomy, LaserLaB, Vrije Universiteit, De Boelelaan 1081, 1081 HV Amsterdam, The Netherlands}

\begin{abstract}
The saturation spectrum of the R(1) transition in the (2-0) band in HD is found to exhibit a composite line shape, involving a Lamb-dip and a Lamb-peak. We propose an explanation for such behavior based on the effects of cross-over resonances in the hyperfine substructure, which is made quantitative in a density-matrix calculation. This resolves an outstanding discrepancy on the rovibrational R(1) transition frequency, which is now determined at 217\,105\,181\,901 (50) kHz and in agreement with current theoretical calculations.
\end{abstract}

\maketitle

Vibrational transitions in the hydrogen deuteride (HD) molecule, associated with the small permanent dipole moment of this hetero-nuclear species, were first detected by Herzberg~\cite{Herzberg1950}. Thereafter transition frequencies of a number of lines in the (2-0) overtone band were spectroscopically investigated early on~\cite{Durie1960,McKellar1974}, and over the decades at increasing accuracy~\cite{Kassi2011}. A vast number of spectroscopic studies have been performed on the HD vibrational spectra, but all were limited by Doppler broadening.
In two recent studies laser precision experiments were performed that, for the first time, saturated the absorption spectrum of the very weak R(1) line at $1.38$ $\mu$m in the (2-0) overtone band in HD and acquired a Doppler-free resonance, resulting in orders-of-magnitude improved accuracies. In one study the method of noise-immune-cavity-enhanced optical heterodyne molecular spectroscopy (NICE-OHMS) was employed~\cite{Cozijn2018}, while in a second study a Lamb-dip was observed via cavity ring-down spectroscopy~\cite{Tao2018}. Where in both experiments the spectroscopy laser was locked to a frequency-comb laser for achieving accuracy at $10^{-10}$ levels, the resulting transition frequencies deviated by 900 kHz, corresponding to $9\sigma$ discrepancy for the combined uncertainties. In view of the importance of such precision measurements for testing quantum electrodynamics in hydrogen molecular systems~\cite{Cheng2018,Puchalski2019},
for probing fifth forces~\cite{Salumbides2013} and extra dimensions~\cite{Salumbides2015b}, we have reinvestigated this R(1) line of HD at improved background suppression and signal-to-noise ratio, and under varying pressure conditions with the aim of finding the cause of this discrepancy.

\begin{figure}[htbp]
\centering
\fbox{\includegraphics[width=0.95\linewidth]{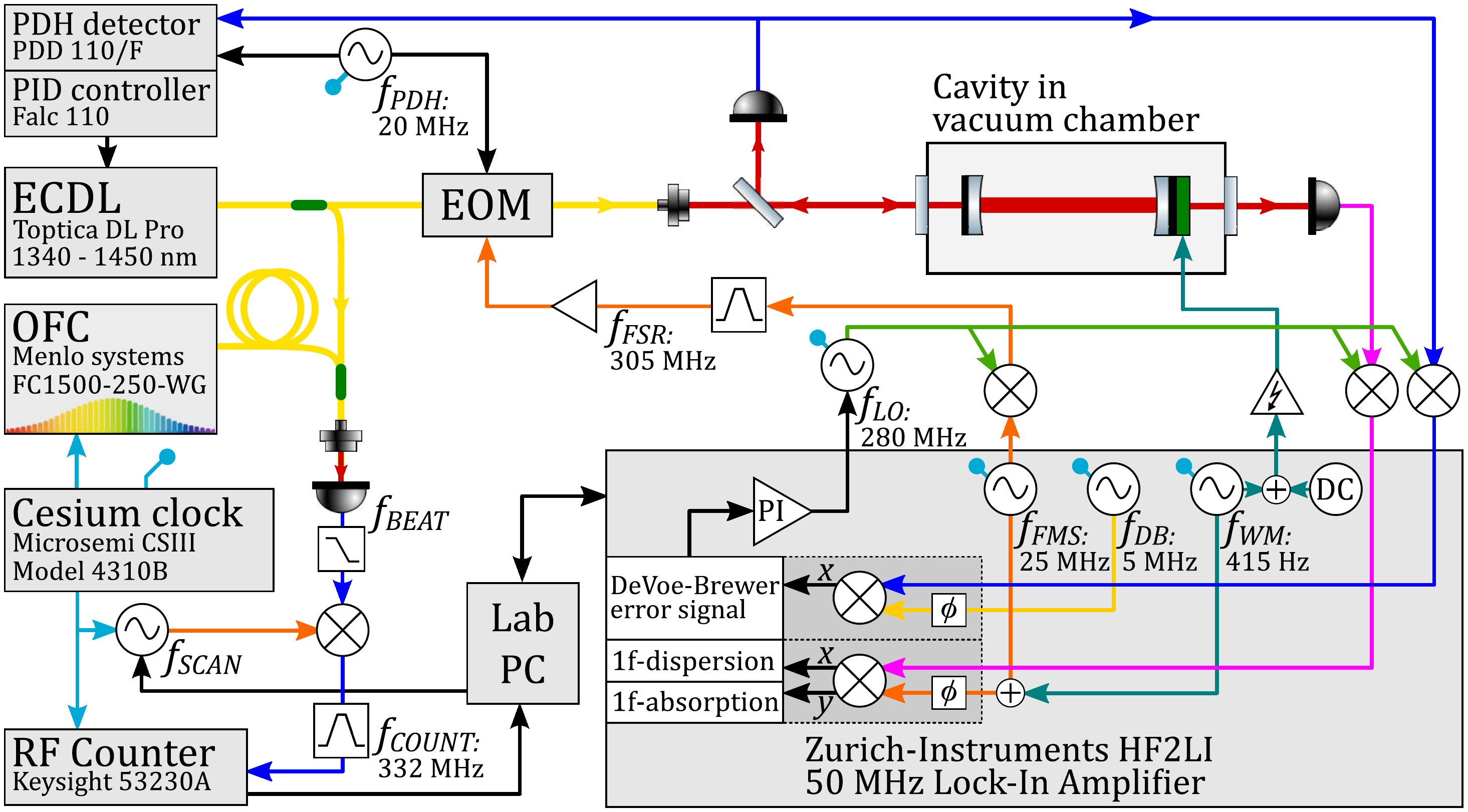}}
\caption{\small{Schematic layout of the experimental setup used in the present study. The ECDL laser is modulated twice by the EOM at both the FSR and PDH frequencies, of which the latter is used for short term stabilization onto the high-finesse cavity. Additionally, the laser is wavelength-modulated through a dither on the cavity piezo for derivative detection. A slow lock with the frequency comb is performed for long term stabilization.}}
\label{fig:setup}
\end{figure}

The frequency-comb locked NICE-OHMS setup described in the previous study~\cite{Cozijn2018} is significantly modified on both the signal and frequency acquisition. It now involves a high-speed lock-in amplifier (Zurich Instruments, HF2LI), which allows parallel demodulation to extract all relevant signal components simultaneously as both the down-converted high-frequency modulation ($f_{FSR}$) and low-frequency dither ($f_{WM}$) all fall inside the lock-in amplifier bandwidth (50 MHz). A schematic layout of the experimental setup is displayed in Fig.~\ref{fig:setup}.

The spectroscopy laser is directly stabilized onto the high-finesse cavity through a Pound-Drever-Hall (PDH) lock for short term stabilization, while long term stability and accuracy ($1$ kHz) is obtained by a direct comparison with a Cs-stabilized frequency comb.
To remove the effect of the periodic frequency dither ($f_{WM}=415$ Hz, $90$ kHz peak-to-peak amplitude) in the acquisition, the counter operates at a gate time set to an integer value of the dither period. The measured beatnote value is used in a digital feedback loop to feedback the cavity piezo.
The DeVoe-Brewer signal, required to stabilize the modulation frequency onto the cavity Free-Spectral-Range $f_{FSR} =305$ MHz, is also retrieved within the lock-in amplifier. It occurs at both frequencies of $f_{FSR} \pm f_{PDH}$, which are located at $5$ MHz or $45$ MHz within the lock-in amplifier due to the down-conversion. To allow retrieval of this error function, it is required that both the internal oscillators of the lock-in amplifier and the PDH frequency generator are all stabilized to an external reference to ensure a fixed phase relationship.

\begin{figure}[htbp]
\centering
\fbox{\includegraphics[width=0.97\linewidth]{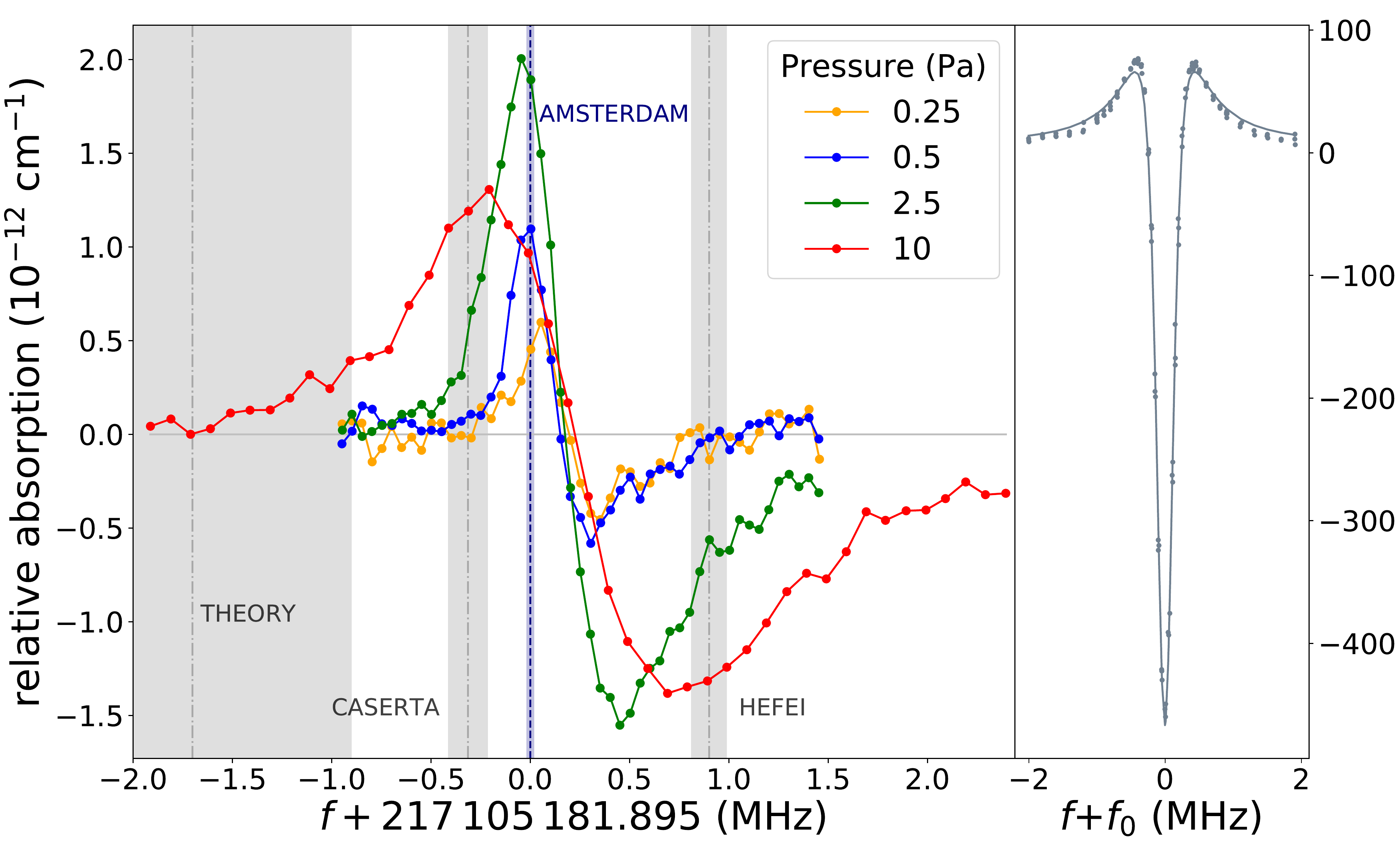}}
\caption{\small{Observed $1f$-spectra of the HD R(1) transition under conditions of varying pressure are shown in the left panel. For comparison in the right panel, an R($9$) acetylene line ($f_0=217\,377\,365\,327$ kHz) recorded at $0.1$ Pa with identical settings of modulation and demodulation phases for the HD measurements.
The acetylene spectrum reflects the expected $1f$-derivative of the dispersion profile for a saturated NICE-OHMS Lamb dip.
The vertical lines in the left panel indicate the line positions reported in Refs.~\cite{Cozijn2018} (Amsterdam), \cite{Tao2018} (Hefei), and \cite{Fasci2018} (Caserta), as well as the theoretical value from Ref.~ \cite{Czachorowski2018}. The shaded areas indicate the uncertainty estimates of the reported line positions.
}}
\label{fig:spec}
\end{figure}

Saturated absorption spectra of the HD R(1) line were measured as a function of pressure in the range 0.25 to 10 Pa, recordings of which are displayed in Fig.~\ref{fig:spec}.
With a cavity finesse of $\sim150\,000$, the circulating power is estimated to be $\sim100$ W.
The spectra represent the average of 8 scans, each taken at 20 minutes recording time (with $50$-kHz frequency steps), yielding more than a five-fold improvement in signal-to-noise ratio with respect to the previous study~\cite{Cozijn2018}.
It should be noted that the NICE-OHMS technique employed, which is essentially a form of frequency modulation spectroscopy in an optical cavity, yields a dispersive spectral line shape~\cite{Axner2014a,Foltynowicz2009b,Twagirayezu2015,Dupre2015a}.
However, the recordings are performed with an additional slow wavelength modulation at frequency $f_{WM}$, which improves the signal-to-noise ratio and allows for detecting the extremely weak HD signal in saturation. Hence, after $1f$ demodulation at $f_{WM}$, a derivative of the NICE-OHMS dispersion channel is expected in the form of a \emph{symmetric} line shape.
Indeed, a symmetric saturated absorption line is observed for C$_2$H$_2$ (an R($9$) line at $271\,377\,365\,327\,(5)$ kHz), recorded under the same NICE-OHMS modulation conditions and displayed in the right panel of Fig.~\ref{fig:spec}.
The C$_2$H$_2$ line reflects the expected line shape, and the characteristic Lamb-dip of saturation spectroscopy.

The measurements performed on the HD R(1) line exhibit a variation in line profiles at different pressures. At low pressures (0.25-1 Pa) the line profile is dominated by a feature with a reversed sign when compared to the C$_2$H$_2$ spectrum. These HD-signals at low pressure form Lamb-peaks, that were interpreted in the previous study~\cite{Cozijn2018} as the saturated line from which the centre frequency of the R(1) line was derived, at $217\,105\,181\,895 (20)$ kHz.

At the lowest pressures ($0.25$ Pa), besides the marked enhanced absorption (peak) at low frequencies some reduced absorption at higher frequencies is observable.
The reduced absorption feature, or a Lamb-dip, becomes more pronounced as a more intense and gradually broader signal with increasing pressure. With these two features, the resonance resembles a dispersive line shape.
This deviation from an expected \emph{symmetric} line shape is the central finding of the present experimental study.

While the Lamb-peak signal at low frequencies corresponds to the feature analyzed in Ref.~\cite{Cozijn2018}, a broader feature, of the same sign as the C$_2$H$_2$ feature is also observed. This feature, having the sign of a Lamb-dip, is located in the range of higher frequencies, where the Lamb-dip of the HD R(1) line was found in the cavity ring-down study, at $217\,105\,182.79 (8)$ MHz ~\cite{Tao2018}. These observations on pressure-dependent line shapes may form the basis for resolving the discrepancy between the two results previously published~\cite{Cozijn2018,Tao2018}.
The line position reported in a recent study, a Doppler-limited measurement using frequency-comb assisted cavity ring-down spectroscopy~\cite{Fasci2018}, is also indicated in Fig.~\ref{fig:spec}.
In the following, an analysis of the line shape is presented that should quantitatively support this assumption, although a number of approximations and hypotheses must be made.

\begin{figure}[htbp]
\centering
\fbox{\includegraphics[width=0.95\linewidth]{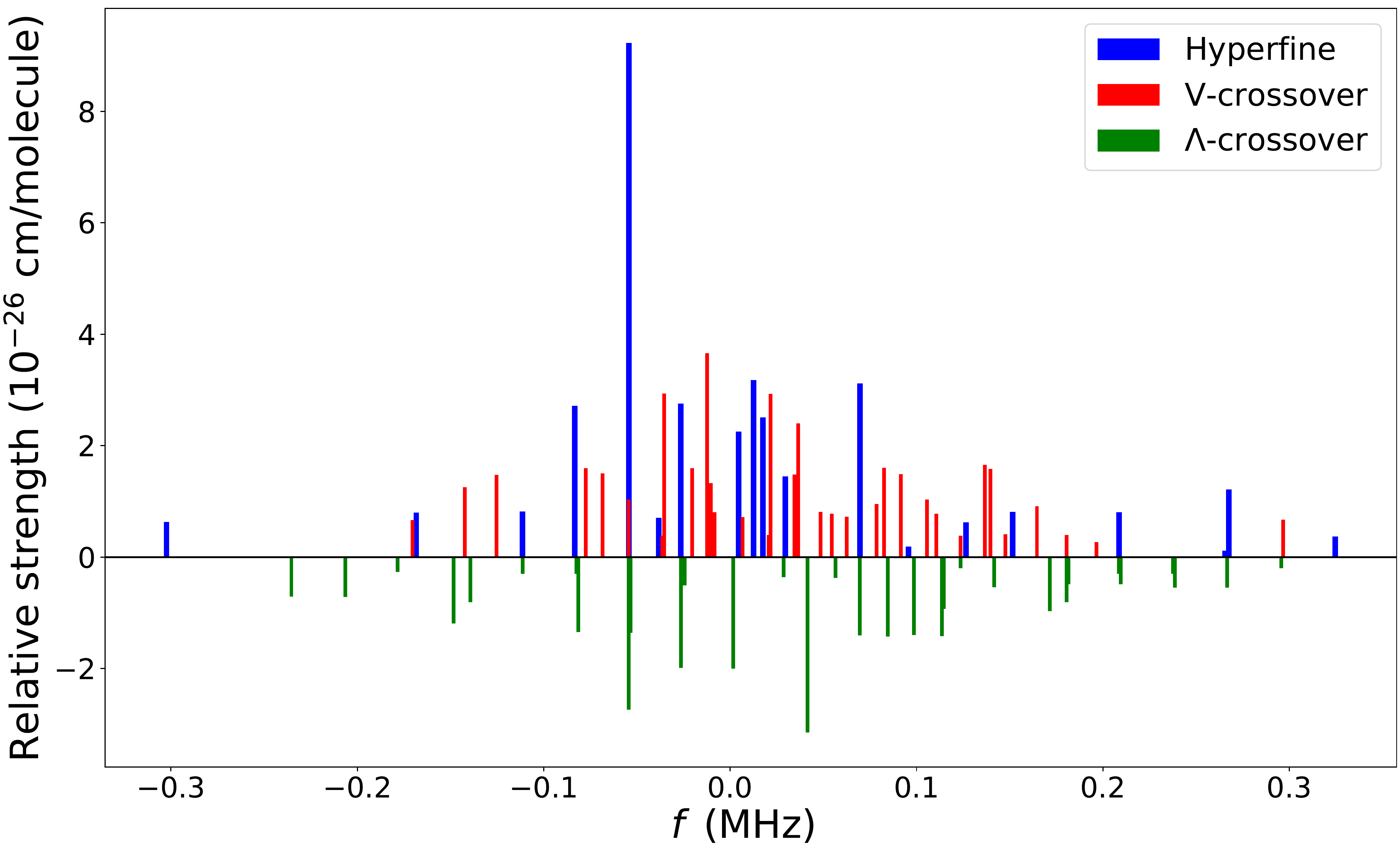}}
\caption{\small{A stick spectrum representing the 21 hyperfine components (in blue) and the 68 cross-over resonances (in red and green) in the R(1) line of HD. The zero on the $x$-axis represents the location of the hyperfineless rovibrational transition. The line strengths of the cross-over resonances represent the root of the product of hyperfine line intensities of the two connecting transitions.}}
\label{fig:HD-R1-sticks}
\end{figure}

Crucial for our proposed interpretation of the line shape is the underlying hyperfine structure of the R(1) transition.
The nuclear spins for the deuteron $I_D=1$ and the proton $I_H=1/2$ give rise to a splitting in 5 hyperfine substates in the $J=1$ ground state and into 6 hyperfine substates in the $J=2$ excited state. This gives rise to 21 possible hyperfine components within the saturated absorption profile of the R(1) line. The level structure for the  $v=0,J=1$ level follows from the analysis of RF-spectra by Ramsey and Lewis~\cite{Ramsey1957}.  For the $v=2$ level hyperfine coupling constants were calculated via \emph{ab initio} theory, and from these the hyperfine level splitting in $v=2,J=2$ were derived and assumed to be accurate within $5\%$ ~\cite{Dupre2019a}. Relative line intensities of the 21 components were calculated via angular momentum algebra~\cite{Dupre2019a}. The calculated hyperfine structure is shown as a stick spectrum in Fig.~\ref{fig:HD-R1-sticks}. This graph also includes the locations of so-called cross-over resonances that couple non-zero velocity classes in a narrow saturation peak. Between the 5 ground state levels and the 6 excited state levels there exist 32 resonances where two ground state levels are connected to a common upper state ($\Lambda$-type cross overs), and 36 resonances that couple two excited levels to a common ground state (V-type cross overs).
The occurrence of cross-over resonances is well documented in saturation spectroscopy~\cite{Foth1979,Bloomfield1982,Hertzler1990,Borde1979}. They give rise to sign reversals, hence Lamb peaks, depending on the V-type or $\Lambda$-type connection of hyperfine levels~\cite{Thomas1980,Schmidt1994}.

To study the general features of the observed HD R(1) line profile, an approximate model was set up using the density matrix formalism that includes the 5 hyperfine sublevels of the $v=0,J=1$ ground level and the 6 hyperfine sublevels of the  $v=2,J=2$ excited level. The coupled Bloch equations involving populations of excited sublevels $\rho_{jj}$ and ground sublevels $\rho_{ii}$ and the coherences $\rho_{ij}$ are:

\begin{align}
    \frac{d}{dt}\rho_{ii}&= \sum_{j}\rho_{jj}\gamma_{pop,ij}-\frac{i}{\hbar}\sum_{j}(V_{ji}\rho_{ji}-\rho_{ji}V_{ij}), \\
    \frac{d}{dt}\rho_{jj}&= -\sum_{i}\rho_{jj}\gamma_{pop,ij}-\frac{i}{\hbar}\sum_{i}(V_{ij}\rho_{ij}-\rho_{ij}V_{ji}),
\end{align}

\begin{multline}
    \frac{d}{dt}\rho_{ij} = -(i\Delta_{ij}+\gamma_{coh,ij})\rho_{ij}+\frac{i}{\hbar}V_{ij}(\rho_{ii}-\rho_{jj})\\
    -\frac{i}{\hbar}\sum_{k\neq j}V_{ik}\rho_{kj},
\end{multline}

\begin{align}
    \frac{d}{dt}\rho_{jk}&= (i\Delta_{jk}-\gamma_{coh,jk})\rho_{jk}-\frac{i}{\hbar}(V_{ij}\rho_{ji}-\rho_{ki}V_{ki}),
\end{align}
with the detuning $\Delta_{ij}=\omega_{L}-(\omega_{ij}+\vec{k}\cdot\vec{v})$, where $\omega_{L}$ is the laser frequency, $\omega_{ij}$ is the transition frequency between states $i$ and $j$, and $\vec{k}\cdot\vec{v}$ is the Doppler shift for the velocity class $v$.
The population relaxation rates $\gamma_{pop,ij}$ connecting $i$ and $j$ states with allowed dipole transitions and relaxation rates for coherences $\gamma_{coh,ij}$ effectively describe radiative and non-radiative processes.
The coupling $V_{ij}=\mu_{ij}(E_{+} + E_{-}) + c.c.$ consists of electric field contributions from forward $+k$ and backward $-k$ propagating laser beams with equal intensities.
The electronic transition dipole moment as well as relative intensities of the hyperfine transitions are included in $\mu_{ij}$. In the calculation, the detuning $\Delta$ of the applied laser field is scanned over the vicinity of the resonances, with the Doppler shift accounted for both $+k$ and $-k$ beams and for each of the $n_v$ velocity classes around $v=0$.
We have neglected the effect of the two laser sidebands at  $f_{FSR}= \pm \, 305$ MHz (at field strengths of about $1\%$ of the carrier), to simplify the treatment.
The occurrence of a recoil doublet~\cite{Bagayev1991}, with a splitting of 68 kHz for the R(1) line, is ignored as well.

Another complication in explaining the observed spectra lies in the reduction of transit-time broadening.
We note here, as was discussed in~\cite{Cozijn2018}, that the observed linewidth of the HD R(1) line is observed much narrower than expected from transit time broadening for a room-temperature velocity distribution. This was attributed to the mechanism of selection of cold molecules under conditions of very weak saturation, which is a known phenomenon~\cite{Bagaev1976,Bagaev1987,Ma1999}. In the model description no attempt was made to explicitly explain this behavior.

The coupled Bloch equations were solved by numerical integration using a Python code and associated numerical and scientific libraries.
The computation was performed in a cluster computer utilizing $32$ nodes with $16$ cores.
For $n_v\sim4400$ velocity classes that cover Doppler shifts in the range of $[-1.8,1.8]$ MHz around each resonance, a calculation of a spectrum typically takes around $18$ hours.

A first calculation was performed that was restricted to the $v=0$ velocity fraction ($v=0$ along the propagating laser beams). Its result shows a spectrum displaying all hyperfine resonances at the expected positions indicated as proper hyperfine resonances in Fig.~\ref{fig:HD-R1-sticks} as Lamb-dips. Subsequently, more elaborate calculations were performed, integrating the spectrum over extended velocity classes. Depending on the parameters invoked for $\gamma_{coh}$ and $\gamma_{pop}$, this treatment produces cross-over resonances with the same sign as a Lamb-dip, but also cross-over features with a reversed-sign. The simulations reveal the effects of velocity selective optical pumping that are observed at particular velocity classes where the cross-over interaction between the counter-propagating beams occur.
From a parameter-space analysis it is found that the sign-reversed cross-over peaks occur only under certain values for the input parameters.
Clear anti-crossovers are produced only when the population decay $\gamma_{pop}$ occurs at a faster rate than the Rabi frequency that was estimated at $V/\hbar=20$ kHz in the measurements~\cite{Cozijn2018}.

This suggests that there must be a mechanism transferring population from the excited $v=2$ state to the lower $v=0$ state in the molecule, a \emph{refilling} mechanism.
This rate should be much larger than the radiative lifetime of the upper HD level, being on the order of $\Gamma\sim$ 1 Hz.
Hence, in this specific case of a saturated transition with a long upper state lifetime in a molecule, similar to the case of NH$_3$~\cite{Lemarchand2011}, the population decay ($\gamma_{pop}$) cannot be induced by spontaneous radiative decay, as in the case of atoms~\cite{Bloomfield1982,Hertzler1990,Schmidt1994}.

In order to effectively produce anti-crossovers or Lamb peaks an hypothesis must be invoked as to the refilling mechanism for ground state population.
In some examples, that bear some similarity, the main contribution to the rate of population decay $\gamma_{pop}$ was attributed to optical pumping in a multi-level system, with the transit-time effect as the absorbers traverse the beam playing a role~\cite{Thomas1980,Schmidt1994}.
In a model for saturation in coherent anti-Stokes spectroscopy in H$_2$ the mechanism of velocity-changing collisions was adopted as a possible transfer mechanism~\cite{Lucht1989}.
The decay value in the latter CARS experiment was close to the value of the pressure broadening coefficient of $\gamma_{P}=35$ kHz/Pa for saturated absorption extracted in Ref.~\cite{Cozijn2018}, while Ref.~\cite{Fasci2018}  reports  a coefficient of $\gamma_{P}=10$ kHz/Pa for a Doppler-limited study (without saturation).
For the pressure $P\sim1$ Pa accessed in the measurements, the collision rate contribution of $\gamma_{P}P$ to $\gamma_{coh,ij}$ is $\sim10-35$ kHz.
Without further speculation on the nature of the mechanism, a range of values for the refilling rate was used in solving the density matrix framework.

\begin{figure}[htbp]
\centering
\fbox{\includegraphics[width=0.95\linewidth]{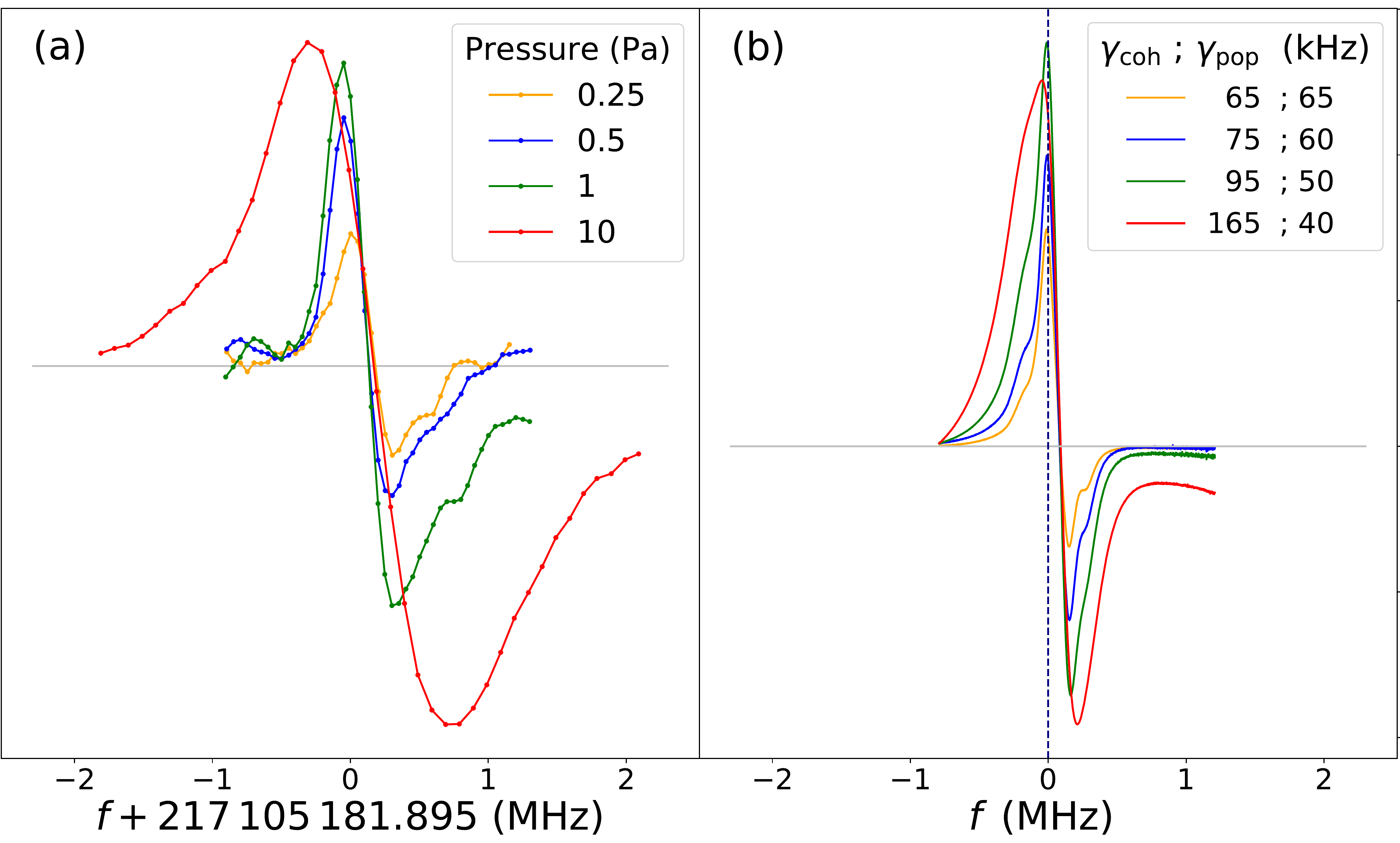}}
\caption{\small{Comparison between observed and modeled spectral line shapes of the R(1) line of HD; (a) Experimental spectra for 0.25, 0.5, 1.0 and 10 Pa on an absolute frequency scale; (b) Calculated spectra on a relative frequency scale with respect to the hyperfineless point (at $0$ MHz) for different values of  $\gamma_{coh}$ and $\gamma_{pop}$ in units of kHz and the Rabi frequency fixed at 20 kHz.}}
\label{fig:Dens-Matrix}
\end{figure}

Figure~\ref{fig:Dens-Matrix} shows numerical results obtained for conditions that mimic the observations of spectra at different pressures.
The line profiles qualitatively agree with experimental observations. In particular the Lamb-peak and dip features correspond to the measurements, although the intensities and widths are not well reproduced. Both the experiments and simulations demonstrate that the Lamb-peak persists at the lowest pressures (lowest values of $\gamma_{coh,ij}$), while the Lamb-dip decreases in intensity at the low pressures.

The simulations show that the saturation line profiles of HD R(1) exhibit a composite line shape with a Lamb-peak at low frequency and a broader Lamb-dip occurring at higher frequencies. We postulate this as an explanation for the results obtained in previous studies, where a low-frequency Lamb-peak was analyzed \cite{Cozijn2018}. We assume that the high-frequency component was measured in the cavity ring down study \cite{Tao2018}, presumably at somewhat higher pressures. We emphasize that a number of approximations were made in the numerical treatment which could affect the intensities and widths, such as the neglect of the recoil doublet, the inclusion of approximate results on the hyperfine structure from calculations, the lack of a quantitative description of the mechanism selecting cold molecules resulting in larger transit times, absence of cavity standing-wave effects, and in particular the lack of understanding of the refilling mechanism.

A comparison can be made between the observed line shapes and the composite profiles from the numerical treatment, i.e. in a fit. Since the numerical analysis is based on the hyperfine components dispersed at well-known frequencies, such a fit results in a value for the hyperfineless HD R(1) rovibrational transition.
It is interesting to note that a pressure-induced shift of $-10$ kHz/Pa  for the Lamb peak line position in the simulations (Fig.~\ref{fig:Dens-Matrix}) is close to that extracted in the experimental results.
By taking into account profiles for various pressures in the simulations, the extrapolated collision-free Lamb peak line center can be related to the hyperfineless transition frequency.
Adopting this treatment to the measurements yields a value of $217\,105\,181\,901$ kHz for the R(1) transition frequency.
In view of the numerous assumptions made and the pressure-dependent line shape we take a conservative estimate for the uncertainty of the hyperfineless transition frequency of some 50 kHz.
The analysis shows also that the zero-pressure extrapolation for the Lamb-peak, as determined in \cite{Cozijn2018}, is rather close to the hyperfine center-of-gravity to within $20$ kHz. The present result agrees, within its $50$ kHz uncertainty, with our previous determination \cite{Cozijn2018}.

The presently obtained transition frequency may be compared to the theoretical values obtained from advanced molecular QED calculations. While a calculation of early 2018 yielded a value of $217\,105\,174.8\,(1.8)$ MHz, included in \cite{Tao2018}, deviating by 7 or 8 MHz from both experimental values, recently a theoretical value was reported at $217\,105\,180.2\,(0.9)$ MHz~\cite{Czachorowski2018}. The latter value, that was based on an improved treatment of relativistic corrections in  non-adiabatic perturbation theory, is within $2\sigma$ from the presently obtained experimental value. This allows for the conclusion that agreement is obtained between experiment and theory of infrared transitions in hydrogen at the $5 \times 10^{-9}$ level, where the present result is of higher accuracy than current state-of-the art calculations. Meanwhile a theoretical framework has been developed that should produce a more accurate value~\cite{Puchalski2019}, to be confronted with the experimental value.
On the experimental side the observation of the R(0) line in the (2-0) band, exhibiting a much simpler hyperfine structure with a near-degenerate ground state~\cite{Puchalski2018}, could be measured. Unfortunately that line is overlaid by a water resonance prohibiting its measurements in the current setup.

\vspace{0.2cm}
The authors thank
J.M.A. Staa (VU) for assisting in the measurements and
P. Dupr\'e (Dijon) for making available calculations on the hyperfine structure.
The work was supported by the Netherlands Organisation for Scientific
Research (NWO) via the program "The mysterious size of the proton".
SURFsara (www.surfsara.nl) is acknowledged for the support in using the Lisa Compute Cluster.
WU acknowledges the European Research Council for an ERC-Advanced Grant under the European Union’s Horizon2020 research and innovation programme (Grant agreement No. 670168).

\bigskip

\bibliography{Hydrogen-WU,NICE-OHMS}

\begin{thebibliography}{31}%
\makeatletter
\providecommand \@ifxundefined [1]{%
 \@ifx{#1\undefined}
}%
\providecommand \@ifnum [1]{%
 \ifnum #1\expandafter \@firstoftwo
 \else \expandafter \@secondoftwo
 \fi
}%
\providecommand \@ifx [1]{%
 \ifx #1\expandafter \@firstoftwo
 \else \expandafter \@secondoftwo
 \fi
}%
\providecommand \natexlab [1]{#1}%
\providecommand \enquote  [1]{``#1''}%
\providecommand \bibnamefont  [1]{#1}%
\providecommand \bibfnamefont [1]{#1}%
\providecommand \citenamefont [1]{#1}%
\providecommand \href@noop [0]{\@secondoftwo}%
\providecommand \href [0]{\begingroup \@sanitize@url \@href}%
\providecommand \@href[1]{\@@startlink{#1}\@@href}%
\providecommand \@@href[1]{\endgroup#1\@@endlink}%
\providecommand \@sanitize@url [0]{\catcode `\\12\catcode `\$12\catcode
  `\&12\catcode `\#12\catcode `\^12\catcode `\_12\catcode `\%12\relax}%
\providecommand \@@startlink[1]{}%
\providecommand \@@endlink[0]{}%
\providecommand \url  [0]{\begingroup\@sanitize@url \@url }%
\providecommand \@url [1]{\endgroup\@href {#1}{\urlprefix }}%
\providecommand \urlprefix  [0]{URL }%
\providecommand \Eprint [0]{\href }%
\providecommand \doibase [0]{http://dx.doi.org/}%
\providecommand \selectlanguage [0]{\@gobble}%
\providecommand \bibinfo  [0]{\@secondoftwo}%
\providecommand \bibfield  [0]{\@secondoftwo}%
\providecommand \translation [1]{[#1]}%
\providecommand \BibitemOpen [0]{}%
\providecommand \bibitemStop [0]{}%
\providecommand \bibitemNoStop [0]{.\EOS\space}%
\providecommand \EOS [0]{\spacefactor3000\relax}%
\providecommand \BibitemShut  [1]{\csname bibitem#1\endcsname}%
\let\auto@bib@innerbib\@empty
\bibitem [{\citenamefont {Herzberg}(1950)}]{Herzberg1950}%
  \BibitemOpen
  \bibfield  {author} {\bibinfo {author} {\bibfnamefont {G.}~\bibnamefont
  {Herzberg}},\ }\href@noop {} {\bibfield  {journal} {\bibinfo  {journal}
  {Nature}\ }\textbf {\bibinfo {volume} {166}},\ \bibinfo {pages} {563}
  (\bibinfo {year} {1950})}\BibitemShut {NoStop}%
\bibitem [{\citenamefont {Durie}\ and\ \citenamefont
  {Herzberg}(1960)}]{Durie1960}%
  \BibitemOpen
  \bibfield  {author} {\bibinfo {author} {\bibfnamefont {R.~A.}\ \bibnamefont
  {Durie}}\ and\ \bibinfo {author} {\bibfnamefont {G.}~\bibnamefont
  {Herzberg}},\ }\href {https://doi.org/10.1139/p60-086} {\bibfield  {journal}
  {\bibinfo  {journal} {Can. J. Phys.}\ }\textbf {\bibinfo {volume} {38}},\
  \bibinfo {pages} {806} (\bibinfo {year} {1960})}\BibitemShut {NoStop}%
\bibitem [{\citenamefont {McKellar}(1974)}]{McKellar1974}%
  \BibitemOpen
  \bibfield  {author} {\bibinfo {author} {\bibfnamefont {A.~R.~W.}\
  \bibnamefont {McKellar}},\ }\href@noop {} {\bibfield  {journal} {\bibinfo
  {journal} {Can. J. Phys.}\ }\textbf {\bibinfo {volume} {52}},\ \bibinfo
  {pages} {1144} (\bibinfo {year} {1974})}\BibitemShut {NoStop}%
\bibitem [{\citenamefont {Kassi}\ and\ \citenamefont
  {Campargue}(2011)}]{Kassi2011}%
  \BibitemOpen
  \bibfield  {author} {\bibinfo {author} {\bibfnamefont {S.}~\bibnamefont
  {Kassi}}\ and\ \bibinfo {author} {\bibfnamefont {A.}~\bibnamefont
  {Campargue}},\ }\href@noop {} {\bibfield  {journal} {\bibinfo  {journal} {J.
  Mol. Spectrosc.}\ }\textbf {\bibinfo {volume} {267}},\ \bibinfo {pages} {36}
  (\bibinfo {year} {2011})}\BibitemShut {NoStop}%
\bibitem [{\citenamefont {Cozijn}\ \emph {et~al.}(2018)\citenamefont {Cozijn},
  \citenamefont {Dupr\'e}, \citenamefont {Salumbides}, \citenamefont {Eikema},\
  and\ \citenamefont {Ubachs}}]{Cozijn2018}%
  \BibitemOpen
  \bibfield  {author} {\bibinfo {author} {\bibfnamefont {F.~M.~J.}\
  \bibnamefont {Cozijn}}, \bibinfo {author} {\bibfnamefont {P.}~\bibnamefont
  {Dupr\'e}}, \bibinfo {author} {\bibfnamefont {E.~J.}\ \bibnamefont
  {Salumbides}}, \bibinfo {author} {\bibfnamefont {K.~S.~E.}\ \bibnamefont
  {Eikema}}, \ and\ \bibinfo {author} {\bibfnamefont {W.}~\bibnamefont
  {Ubachs}},\ }\href@noop {} {\bibfield  {journal} {\bibinfo  {journal} {Phys.
  Rev. Lett.}\ }\textbf {\bibinfo {volume} {120}},\ \bibinfo {pages} {153002}
  (\bibinfo {year} {2018})}\BibitemShut {NoStop}%
\bibitem [{\citenamefont {Tao}\ \emph {et~al.}(2018)\citenamefont {Tao},
  \citenamefont {Liu}, \citenamefont {Pachucki}, \citenamefont {Komasa},
  \citenamefont {Sun}, \citenamefont {Wang},\ and\ \citenamefont
  {Hu}}]{Tao2018}%
  \BibitemOpen
  \bibfield  {author} {\bibinfo {author} {\bibfnamefont {L.-G.}\ \bibnamefont
  {Tao}}, \bibinfo {author} {\bibfnamefont {A.-W.}\ \bibnamefont {Liu}},
  \bibinfo {author} {\bibfnamefont {K.}~\bibnamefont {Pachucki}}, \bibinfo
  {author} {\bibfnamefont {J.}~\bibnamefont {Komasa}}, \bibinfo {author}
  {\bibfnamefont {Y.~R.}\ \bibnamefont {Sun}}, \bibinfo {author} {\bibfnamefont
  {J.}~\bibnamefont {Wang}}, \ and\ \bibinfo {author} {\bibfnamefont {S.-M.}\
  \bibnamefont {Hu}},\ }\href@noop {} {\bibfield  {journal} {\bibinfo
  {journal} {Phys. Rev. Lett.}\ }\textbf {\bibinfo {volume} {120}},\ \bibinfo
  {pages} {153001} (\bibinfo {year} {2018})}\BibitemShut {NoStop}%
\bibitem [{\citenamefont {Cheng}\ \emph {et~al.}(2018)\citenamefont {Cheng},
  \citenamefont {Hussels}, \citenamefont {Niu}, \citenamefont {Bethlem},
  \citenamefont {Eikema}, \citenamefont {Salumbides}, \citenamefont {Ubachs},
  \citenamefont {Beyer}, \citenamefont {H\"olsch}, \citenamefont {Agner},
  \citenamefont {Merkt}, \citenamefont {Tao}, \citenamefont {Hu},\ and\
  \citenamefont {Jungen}}]{Cheng2018}%
  \BibitemOpen
  \bibfield  {author} {\bibinfo {author} {\bibfnamefont {C.-F.}\ \bibnamefont
  {Cheng}}, \bibinfo {author} {\bibfnamefont {J.}~\bibnamefont {Hussels}},
  \bibinfo {author} {\bibfnamefont {M.}~\bibnamefont {Niu}}, \bibinfo {author}
  {\bibfnamefont {H.~L.}\ \bibnamefont {Bethlem}}, \bibinfo {author}
  {\bibfnamefont {K.~S.~E.}\ \bibnamefont {Eikema}}, \bibinfo {author}
  {\bibfnamefont {E.~J.}\ \bibnamefont {Salumbides}}, \bibinfo {author}
  {\bibfnamefont {W.}~\bibnamefont {Ubachs}}, \bibinfo {author} {\bibfnamefont
  {M.}~\bibnamefont {Beyer}}, \bibinfo {author} {\bibfnamefont
  {N.}~\bibnamefont {H\"olsch}}, \bibinfo {author} {\bibfnamefont {J.~A.}\
  \bibnamefont {Agner}}, \bibinfo {author} {\bibfnamefont {F.}~\bibnamefont
  {Merkt}}, \bibinfo {author} {\bibfnamefont {L.-G.}\ \bibnamefont {Tao}},
  \bibinfo {author} {\bibfnamefont {S.-M.}\ \bibnamefont {Hu}}, \ and\ \bibinfo
  {author} {\bibfnamefont {C.}~\bibnamefont {Jungen}},\ }\href {\doibase
  10.1103/PhysRevLett.121.013001} {\bibfield  {journal} {\bibinfo  {journal}
  {Phys. Rev. Lett.}\ }\textbf {\bibinfo {volume} {121}},\ \bibinfo {pages}
  {013001} (\bibinfo {year} {2018})}\BibitemShut {NoStop}%
\bibitem [{\citenamefont {Puchalski}\ \emph {et~al.}(2019)\citenamefont
  {Puchalski}, \citenamefont {Komasa}, \citenamefont {Czachorowski},\ and\
  \citenamefont {Pachucki}}]{Puchalski2019}%
  \BibitemOpen
  \bibfield  {author} {\bibinfo {author} {\bibfnamefont {M.}~\bibnamefont
  {Puchalski}}, \bibinfo {author} {\bibfnamefont {J.}~\bibnamefont {Komasa}},
  \bibinfo {author} {\bibfnamefont {P.}~\bibnamefont {Czachorowski}}, \ and\
  \bibinfo {author} {\bibfnamefont {K.}~\bibnamefont {Pachucki}},\ }\href@noop
  {} {\bibfield  {journal} {\bibinfo  {journal} {Phys. Rev. Lett.}\ }\textbf
  {\bibinfo {volume} {122}},\ \bibinfo {pages} {103003} (\bibinfo {year}
  {2019})}\BibitemShut {NoStop}%
\bibitem [{\citenamefont {Salumbides}\ \emph {et~al.}(2013)\citenamefont
  {Salumbides}, \citenamefont {Koelemeij}, \citenamefont {Komasa},
  \citenamefont {Pachucki}, \citenamefont {Eikema},\ and\ \citenamefont
  {Ubachs}}]{Salumbides2013}%
  \BibitemOpen
  \bibfield  {author} {\bibinfo {author} {\bibfnamefont {E.~J.}\ \bibnamefont
  {Salumbides}}, \bibinfo {author} {\bibfnamefont {J.~C.~J.}\ \bibnamefont
  {Koelemeij}}, \bibinfo {author} {\bibfnamefont {J.}~\bibnamefont {Komasa}},
  \bibinfo {author} {\bibfnamefont {K.}~\bibnamefont {Pachucki}}, \bibinfo
  {author} {\bibfnamefont {K.~S.~E.}\ \bibnamefont {Eikema}}, \ and\ \bibinfo
  {author} {\bibfnamefont {W.}~\bibnamefont {Ubachs}},\ }\href@noop {}
  {\bibfield  {journal} {\bibinfo  {journal} {Phys. Rev. D}\ }\textbf {\bibinfo
  {volume} {87}},\ \bibinfo {pages} {112008} (\bibinfo {year}
  {2013})}\BibitemShut {NoStop}%
\bibitem [{\citenamefont {Salumbides}\ \emph {et~al.}(2015)\citenamefont
  {Salumbides}, \citenamefont {Schellekens}, \citenamefont {Gato-Rivera},\ and\
  \citenamefont {Ubachs}}]{Salumbides2015b}%
  \BibitemOpen
  \bibfield  {author} {\bibinfo {author} {\bibfnamefont {E.~J.}\ \bibnamefont
  {Salumbides}}, \bibinfo {author} {\bibfnamefont {A.~N.}\ \bibnamefont
  {Schellekens}}, \bibinfo {author} {\bibfnamefont {B.}~\bibnamefont
  {Gato-Rivera}}, \ and\ \bibinfo {author} {\bibfnamefont {W.}~\bibnamefont
  {Ubachs}},\ }\href@noop {} {\bibfield  {journal} {\bibinfo  {journal} {New J.
  Phys.}\ }\textbf {\bibinfo {volume} {17}},\ \bibinfo {pages} {033015}
  (\bibinfo {year} {2015})}\BibitemShut {NoStop}%
\bibitem [{\citenamefont {Axner}\ \emph {et~al.}(2014)\citenamefont {Axner},
  \citenamefont {Ehlers}, \citenamefont {Foltynowicz}, \citenamefont
  {Silander},\ and\ \citenamefont {Wang}}]{Axner2014a}%
  \BibitemOpen
  \bibfield  {author} {\bibinfo {author} {\bibfnamefont {O.}~\bibnamefont
  {Axner}}, \bibinfo {author} {\bibfnamefont {P.}~\bibnamefont {Ehlers}},
  \bibinfo {author} {\bibfnamefont {A.}~\bibnamefont {Foltynowicz}}, \bibinfo
  {author} {\bibfnamefont {I.}~\bibnamefont {Silander}}, \ and\ \bibinfo
  {author} {\bibfnamefont {J.}~\bibnamefont {Wang}},\ }\enquote {\bibinfo
  {title} {{NICE-OHMS}-frequency modulation cavity-enhanced
  spectroscopy--principles and performance},}\ in\ \href {\doibase
  10.1007/978-3-642-40003-2_6} {\emph {\bibinfo {booktitle} {Cavity-Enhanced
  Spectroscopy and Sensing}}},\ \bibinfo {series and number} {Springer Series
  in Optical Sciences 179}\ (\bibinfo  {publisher} {Springer},\ \bibinfo {year}
  {2014})\ Chap.~\bibinfo {chapter} {6}, pp.\ \bibinfo {pages}
  {211--251}\BibitemShut {NoStop}%
\bibitem [{\citenamefont {Foltynowicz}\ \emph {et~al.}(2009)\citenamefont
  {Foltynowicz}, \citenamefont {Ma}, \citenamefont {Schmidt},\ and\
  \citenamefont {Axner}}]{Foltynowicz2009b}%
  \BibitemOpen
  \bibfield  {author} {\bibinfo {author} {\bibfnamefont {A.}~\bibnamefont
  {Foltynowicz}}, \bibinfo {author} {\bibfnamefont {W.}~\bibnamefont {Ma}},
  \bibinfo {author} {\bibfnamefont {F.}~\bibnamefont {Schmidt}}, \ and\
  \bibinfo {author} {\bibfnamefont {O.}~\bibnamefont {Axner}},\ }\href
  {\doibase 10.1364/JOSAB.26.001384} {\bibfield  {journal} {\bibinfo  {journal}
  {J. Opt. Soc. Am. B}\ }\textbf {\bibinfo {volume} {26}},\ \bibinfo {pages}
  {1384} (\bibinfo {year} {2009})}\BibitemShut {NoStop}%
\bibitem [{\citenamefont {Twagirayezu}\ \emph {et~al.}(2015)\citenamefont
  {Twagirayezu}, \citenamefont {Cich}, \citenamefont {Sears}, \citenamefont
  {McRaven},\ and\ \citenamefont {Hall}}]{Twagirayezu2015}%
  \BibitemOpen
  \bibfield  {author} {\bibinfo {author} {\bibfnamefont {S.}~\bibnamefont
  {Twagirayezu}}, \bibinfo {author} {\bibfnamefont {M.~J.}\ \bibnamefont
  {Cich}}, \bibinfo {author} {\bibfnamefont {T.~J.}\ \bibnamefont {Sears}},
  \bibinfo {author} {\bibfnamefont {C.~P.}\ \bibnamefont {McRaven}}, \ and\
  \bibinfo {author} {\bibfnamefont {G.~E.}\ \bibnamefont {Hall}},\ }\href
  {\doibase https://doi.org/10.1016/j.jms.2015.06.010} {\bibfield  {journal}
  {\bibinfo  {journal} {J. Mol. Spectr.}\ }\textbf {\bibinfo {volume} {316}},\
  \bibinfo {pages} {64 } (\bibinfo {year} {2015})}\BibitemShut {NoStop}%
\bibitem [{\citenamefont {Dupr\'e}(2015)}]{Dupre2015a}%
  \BibitemOpen
  \bibfield  {author} {\bibinfo {author} {\bibfnamefont {P.}~\bibnamefont
  {Dupr\'e}},\ }\href {\doibase 10.1364/JOSAB.32.000838} {\bibfield  {journal}
  {\bibinfo  {journal} {J. Opt. Soc. Am. B}\ }\textbf {\bibinfo {volume}
  {32}},\ \bibinfo {pages} {838} (\bibinfo {year} {2015})}\BibitemShut
  {NoStop}%
\bibitem [{\citenamefont {Fasci}\ \emph {et~al.}(2018)\citenamefont {Fasci},
  \citenamefont {Castrillo}, \citenamefont {Dinesan}, \citenamefont {Gravina},
  \citenamefont {Moretti},\ and\ \citenamefont {Gianfrani}}]{Fasci2018}%
  \BibitemOpen
  \bibfield  {author} {\bibinfo {author} {\bibfnamefont {E.}~\bibnamefont
  {Fasci}}, \bibinfo {author} {\bibfnamefont {A.}~\bibnamefont {Castrillo}},
  \bibinfo {author} {\bibfnamefont {H.}~\bibnamefont {Dinesan}}, \bibinfo
  {author} {\bibfnamefont {S.}~\bibnamefont {Gravina}}, \bibinfo {author}
  {\bibfnamefont {L.}~\bibnamefont {Moretti}}, \ and\ \bibinfo {author}
  {\bibfnamefont {L.}~\bibnamefont {Gianfrani}},\ }\href {\doibase
  10.1103/PhysRevA.98.022516} {\bibfield  {journal} {\bibinfo  {journal} {Phys.
  Rev. A}\ }\textbf {\bibinfo {volume} {98}},\ \bibinfo {pages} {022516}
  (\bibinfo {year} {2018})}\BibitemShut {NoStop}%
\bibitem [{\citenamefont {Czachorowski}\ \emph {et~al.}(2018)\citenamefont
  {Czachorowski}, \citenamefont {Puchalski}, \citenamefont {Komasa},\ and\
  \citenamefont {Pachucki}}]{Czachorowski2018}%
  \BibitemOpen
  \bibfield  {author} {\bibinfo {author} {\bibfnamefont {P.}~\bibnamefont
  {Czachorowski}}, \bibinfo {author} {\bibfnamefont {M.}~\bibnamefont
  {Puchalski}}, \bibinfo {author} {\bibfnamefont {J.}~\bibnamefont {Komasa}}, \
  and\ \bibinfo {author} {\bibfnamefont {K.}~\bibnamefont {Pachucki}},\
  }\href@noop {} {\bibfield  {journal} {\bibinfo  {journal} {Phys. Rev. A}\
  }\textbf {\bibinfo {volume} {98}},\ \bibinfo {pages} {052506} (\bibinfo
  {year} {2018})}\BibitemShut {NoStop}%
\bibitem [{\citenamefont {Ramsey}\ and\ \citenamefont
  {Lewis}(1957)}]{Ramsey1957}%
  \BibitemOpen
  \bibfield  {author} {\bibinfo {author} {\bibfnamefont {N.~F.}\ \bibnamefont
  {Ramsey}}\ and\ \bibinfo {author} {\bibfnamefont {H.~R.}\ \bibnamefont
  {Lewis}},\ }\href@noop {} {\bibfield  {journal} {\bibinfo  {journal} {Phys.
  Rev.}\ }\textbf {\bibinfo {volume} {108}},\ \bibinfo {pages} {1246} (\bibinfo
  {year} {1957})}\BibitemShut {NoStop}%
\bibitem [{\citenamefont {Dupr\'e}\ and\ \citenamefont
  {Gauss}(2019)}]{Dupre2019a}%
  \BibitemOpen
  \bibfield  {author} {\bibinfo {author} {\bibfnamefont {P.}~\bibnamefont
  {Dupr\'e}}\ and\ \bibinfo {author} {\bibfnamefont {J.}~\bibnamefont
  {Gauss}},\ }\href@noop {} {\bibfield  {journal} {\bibinfo  {journal} {private
  communication}\ } (\bibinfo {year} {2019})}\BibitemShut {NoStop}%
\bibitem [{\citenamefont {Foth}\ and\ \citenamefont
  {Spieweck}(1979)}]{Foth1979}%
  \BibitemOpen
  \bibfield  {author} {\bibinfo {author} {\bibfnamefont {H.}~\bibnamefont
  {Foth}}\ and\ \bibinfo {author} {\bibfnamefont {F.}~\bibnamefont
  {Spieweck}},\ }\href {\doibase https://doi.org/10.1016/0009-2614(79)87079-7}
  {\bibfield  {journal} {\bibinfo  {journal} {Chem. Phys. Lett.}\ }\textbf
  {\bibinfo {volume} {65}},\ \bibinfo {pages} {347 } (\bibinfo {year}
  {1979})}\BibitemShut {NoStop}%
\bibitem [{\citenamefont {Bloomfield}\ \emph {et~al.}(1982)\citenamefont
  {Bloomfield}, \citenamefont {Gerhardt}, \citenamefont {Hansch},\ and\
  \citenamefont {Rand}}]{Bloomfield1982}%
  \BibitemOpen
  \bibfield  {author} {\bibinfo {author} {\bibfnamefont {L.~A.}\ \bibnamefont
  {Bloomfield}}, \bibinfo {author} {\bibfnamefont {H.}~\bibnamefont
  {Gerhardt}}, \bibinfo {author} {\bibfnamefont {T.~W.}\ \bibnamefont
  {Hansch}}, \ and\ \bibinfo {author} {\bibfnamefont {S.~C.}\ \bibnamefont
  {Rand}},\ }\href {\doibase https://doi.org/10.1016/0030-4018(82)90026-8}
  {\bibfield  {journal} {\bibinfo  {journal} {Opt. Comm.}\ }\textbf {\bibinfo
  {volume} {42}},\ \bibinfo {pages} {247 } (\bibinfo {year}
  {1982})}\BibitemShut {NoStop}%
\bibitem [{\citenamefont {Hertzler}\ and\ \citenamefont
  {Foth}(1990)}]{Hertzler1990}%
  \BibitemOpen
  \bibfield  {author} {\bibinfo {author} {\bibfnamefont {C.}~\bibnamefont
  {Hertzler}}\ and\ \bibinfo {author} {\bibfnamefont {H.-J.}\ \bibnamefont
  {Foth}},\ }\href {\doibase https://doi.org/10.1016/0009-2614(90)87150-P}
  {\bibfield  {journal} {\bibinfo  {journal} {Chem. Phys. Lett.}\ }\textbf
  {\bibinfo {volume} {166}},\ \bibinfo {pages} {551 } (\bibinfo {year}
  {1990})}\BibitemShut {NoStop}%
\bibitem [{\citenamefont {Bord\'e}\ and\ \citenamefont
  {Bord\'e}(1979)}]{Borde1979}%
  \BibitemOpen
  \bibfield  {author} {\bibinfo {author} {\bibfnamefont {J.}~\bibnamefont
  {Bord\'e}}\ and\ \bibinfo {author} {\bibfnamefont {C.}~\bibnamefont
  {Bord\'e}},\ }\href {\doibase https://doi.org/10.1016/0022-2852(79)90063-8}
  {\bibfield  {journal} {\bibinfo  {journal} {J. Mol. Spectr.}\ }\textbf
  {\bibinfo {volume} {78}},\ \bibinfo {pages} {353 } (\bibinfo {year}
  {1979})}\BibitemShut {NoStop}%
\bibitem [{\citenamefont {Thomas}\ and\ \citenamefont
  {Quivers}(1980)}]{Thomas1980}%
  \BibitemOpen
  \bibfield  {author} {\bibinfo {author} {\bibfnamefont {J.~E.}\ \bibnamefont
  {Thomas}}\ and\ \bibinfo {author} {\bibfnamefont {W.~W.}\ \bibnamefont
  {Quivers}},\ }\href {\doibase 10.1103/PhysRevA.22.2115} {\bibfield  {journal}
  {\bibinfo  {journal} {Phys. Rev. A}\ }\textbf {\bibinfo {volume} {22}},\
  \bibinfo {pages} {2115} (\bibinfo {year} {1980})}\BibitemShut {NoStop}%
\bibitem [{\citenamefont {Schmidt}\ \emph {et~al.}(1994)\citenamefont
  {Schmidt}, \citenamefont {Knaak}, \citenamefont {Wynands},\ and\
  \citenamefont {Meschede}}]{Schmidt1994}%
  \BibitemOpen
  \bibfield  {author} {\bibinfo {author} {\bibfnamefont {O.}~\bibnamefont
  {Schmidt}}, \bibinfo {author} {\bibfnamefont {K.~M.}\ \bibnamefont {Knaak}},
  \bibinfo {author} {\bibfnamefont {R.}~\bibnamefont {Wynands}}, \ and\
  \bibinfo {author} {\bibfnamefont {D.}~\bibnamefont {Meschede}},\ }\href
  {\doibase 10.1007/BF01081167} {\bibfield  {journal} {\bibinfo  {journal}
  {Applied Physics B}\ }\textbf {\bibinfo {volume} {59}},\ \bibinfo {pages}
  {167} (\bibinfo {year} {1994})}\BibitemShut {NoStop}%
\bibitem [{\citenamefont {Bagayev}\ \emph {et~al.}(1991)\citenamefont
  {Bagayev}, \citenamefont {Chebotayev}, \citenamefont {Dmitriyev},
  \citenamefont {Om}, \citenamefont {Nekrasov},\ and\ \citenamefont
  {Skvortsov}}]{Bagayev1991}%
  \BibitemOpen
  \bibfield  {author} {\bibinfo {author} {\bibfnamefont {S.~N.}\ \bibnamefont
  {Bagayev}}, \bibinfo {author} {\bibfnamefont {V.~P.}\ \bibnamefont
  {Chebotayev}}, \bibinfo {author} {\bibfnamefont {A.~K.}\ \bibnamefont
  {Dmitriyev}}, \bibinfo {author} {\bibfnamefont {A.~E.}\ \bibnamefont {Om}},
  \bibinfo {author} {\bibfnamefont {Y.~V.}\ \bibnamefont {Nekrasov}}, \ and\
  \bibinfo {author} {\bibfnamefont {B.~N.}\ \bibnamefont {Skvortsov}},\ }\href
  {\doibase 10.1007/BF00405688} {\bibfield  {journal} {\bibinfo  {journal}
  {Appl. Phys. B}\ }\textbf {\bibinfo {volume} {52}},\ \bibinfo {pages} {63}
  (\bibinfo {year} {1991})}\BibitemShut {NoStop}%
\bibitem [{\citenamefont {Bagaev}\ \emph {et~al.}(1976)\citenamefont {Bagaev},
  \citenamefont {Vasilenko}, \citenamefont {Dmitriev}, \citenamefont
  {Skvortsov},\ and\ \citenamefont {Chebotaev}}]{Bagaev1976}%
  \BibitemOpen
  \bibfield  {author} {\bibinfo {author} {\bibfnamefont {S.~N.}\ \bibnamefont
  {Bagaev}}, \bibinfo {author} {\bibfnamefont {L.~S.}\ \bibnamefont
  {Vasilenko}}, \bibinfo {author} {\bibfnamefont {A.~K.}\ \bibnamefont
  {Dmitriev}}, \bibinfo {author} {\bibfnamefont {M.~N.}\ \bibnamefont
  {Skvortsov}}, \ and\ \bibinfo {author} {\bibfnamefont {V.~P.}\ \bibnamefont
  {Chebotaev}},\ }\href@noop {} {\bibfield  {journal} {\bibinfo  {journal}
  {JETP Lett.}\ }\textbf {\bibinfo {volume} {23}},\ \bibinfo {pages} {360}
  (\bibinfo {year} {1976})}\BibitemShut {NoStop}%
\bibitem [{\citenamefont {Bagaev}\ \emph {et~al.}(1987)\citenamefont {Bagaev},
  \citenamefont {Baklanov}, \citenamefont {Dychkov}, \citenamefont {Pokasov},\
  and\ \citenamefont {Chebotaev}}]{Bagaev1987}%
  \BibitemOpen
  \bibfield  {author} {\bibinfo {author} {\bibfnamefont {S.~N.}\ \bibnamefont
  {Bagaev}}, \bibinfo {author} {\bibfnamefont {A.~E.}\ \bibnamefont
  {Baklanov}}, \bibinfo {author} {\bibfnamefont {A.~S.}\ \bibnamefont
  {Dychkov}}, \bibinfo {author} {\bibfnamefont {P.~V.}\ \bibnamefont
  {Pokasov}}, \ and\ \bibinfo {author} {\bibfnamefont {V.~P.}\ \bibnamefont
  {Chebotaev}},\ }\href@noop {} {\bibfield  {journal} {\bibinfo  {journal}
  {JETP Lett.}\ }\textbf {\bibinfo {volume} {45}},\ \bibinfo {pages} {471}
  (\bibinfo {year} {1987})}\BibitemShut {NoStop}%
\bibitem [{\citenamefont {Ma}\ \emph {et~al.}(1999)\citenamefont {Ma},
  \citenamefont {Ye}, \citenamefont {Dub\'e},\ and\ \citenamefont
  {Hall}}]{Ma1999}%
  \BibitemOpen
  \bibfield  {author} {\bibinfo {author} {\bibfnamefont {L.-S.}\ \bibnamefont
  {Ma}}, \bibinfo {author} {\bibfnamefont {J.}~\bibnamefont {Ye}}, \bibinfo
  {author} {\bibfnamefont {P.}~\bibnamefont {Dub\'e}}, \ and\ \bibinfo {author}
  {\bibfnamefont {J.}~\bibnamefont {Hall}},\ }\href@noop {} {\bibfield
  {journal} {\bibinfo  {journal} {J. Opt. Soc. Am. B}\ }\textbf {\bibinfo
  {volume} {16}},\ \bibinfo {pages} {2255} (\bibinfo {year}
  {1999})}\BibitemShut {NoStop}%
\bibitem [{\citenamefont {Lemarchand}\ \emph {et~al.}(2011)\citenamefont
  {Lemarchand}, \citenamefont {Triki}, \citenamefont {Darqui{\'{e}}},
  \citenamefont {Bord{\'{e}}}, \citenamefont {Chardonnet},\ and\ \citenamefont
  {Daussy}}]{Lemarchand2011}%
  \BibitemOpen
  \bibfield  {author} {\bibinfo {author} {\bibfnamefont {C.}~\bibnamefont
  {Lemarchand}}, \bibinfo {author} {\bibfnamefont {M.}~\bibnamefont {Triki}},
  \bibinfo {author} {\bibfnamefont {B.}~\bibnamefont {Darqui{\'{e}}}}, \bibinfo
  {author} {\bibfnamefont {C.~J.}\ \bibnamefont {Bord{\'{e}}}}, \bibinfo
  {author} {\bibfnamefont {C.}~\bibnamefont {Chardonnet}}, \ and\ \bibinfo
  {author} {\bibfnamefont {C.}~\bibnamefont {Daussy}},\ }\href {\doibase
  10.1088/1367-2630/13/7/073028} {\bibfield  {journal} {\bibinfo  {journal}
  {New J. Phys.}\ }\textbf {\bibinfo {volume} {13}},\ \bibinfo {pages} {073028}
  (\bibinfo {year} {2011})}\BibitemShut {NoStop}%
\bibitem [{\citenamefont {Lucht}\ and\ \citenamefont
  {Farrow}(1989)}]{Lucht1989}%
  \BibitemOpen
  \bibfield  {author} {\bibinfo {author} {\bibfnamefont {R.~P.}\ \bibnamefont
  {Lucht}}\ and\ \bibinfo {author} {\bibfnamefont {R.~L.}\ \bibnamefont
  {Farrow}},\ }\href {\doibase 10.1364/JOSAB.6.002313} {\bibfield  {journal}
  {\bibinfo  {journal} {J. Opt. Soc. Am. B}\ }\textbf {\bibinfo {volume} {6}},\
  \bibinfo {pages} {2313} (\bibinfo {year} {1989})}\BibitemShut {NoStop}%
\bibitem [{\citenamefont {Puchalski}\ \emph {et~al.}(2018)\citenamefont
  {Puchalski}, \citenamefont {Komasa},\ and\ \citenamefont
  {Pachucki}}]{Puchalski2018}%
  \BibitemOpen
  \bibfield  {author} {\bibinfo {author} {\bibfnamefont {M.}~\bibnamefont
  {Puchalski}}, \bibinfo {author} {\bibfnamefont {J.}~\bibnamefont {Komasa}}, \
  and\ \bibinfo {author} {\bibfnamefont {K.}~\bibnamefont {Pachucki}},\
  }\href@noop {} {\bibfield  {journal} {\bibinfo  {journal} {Phys. Rev. Lett.}\
  }\textbf {\bibinfo {volume} {120}},\ \bibinfo {pages} {083001} (\bibinfo
  {year} {2018})}\BibitemShut {NoStop}%
\end{thebibliography}%

\end{document}